\documentclass[12pt,preprint]{aastex6}

\def\apj{{ApJ}}
\def\apjl{{ApJL}}
\def\mnras{{ MNRAS}}

\def\be{\begin{equation}}
\def\ee{\end{equation}}
\def\bea{\begin{eqnarray}}
\def\eea{\end{eqnarray}}

\usepackage{color,txfonts}
\usepackage{hyperref}
\usepackage{epstopdf}
\usepackage{graphicx}
\usepackage[FIGTOPCAP]{subfigure}
\usepackage{CJKutf8}

\begin{document}
\title{Is GW190425 consistent with being a neutron star$-$black hole merger?}
\author{Ming-Zhe Han$^{1,2}$, Shao-Peng Tang$^{1,2}$, Yi-Ming Hu$^{3}$, Yin-Jie Li$^{1,2}$, Jin-Liang Jiang$^{1,2}$, Zhi-Ping Jin$^{1,2}$, Yi-Zhong Fan$^{1,2}$, and Da-Ming Wei$^{1,2}$}
\affiliation{
$^1$ {Key Laboratory of Dark Matter and Space Astronomy, Purple Mountain Observatory, Chinese Academy of Sciences, Nanjing 210034, China }\\
$^2$ {School of Astronomy and Space Science, University of Science and Technology of China, Hefei, Anhui 230026, China.}\\
$^3$ {TianQin Research Center for Gravitational Physics \& School of Physics and Astronomy, Sun Yat-sen University, 2 Daxue Road, Zhuhai 519082, China}\\
}
\email{The corresponding author: yzfan@pmo.ac.cn (Y.Z.F)}

\begin{abstract}
GW190425 is the second neutron star merger event detected by the Advanced LIGO/Virgo detectors. If interpreted as a double neutron star merger, the total gravitational mass is substantially larger than that of the binary systems identified in the Galaxy. In this work we analyze the gravitational-wave data within the neutron star$-$black hole merger scenario. For the black hole, we yield a mass of $2.40^{+0.36}_{-0.32}M_\odot$ and an aligned spin of $0.141^{+0.067}_{-0.064}$. As for the neutron star we find a mass of $1.15^{+0.15}_{-0.13}M_\odot$ and the dimensionless tidal deformability of $1.4^{+3.8}_{-1.2}\times 10^{3}$. These parameter ranges are for 90\% credibility. The inferred masses of the neutron star and the black hole are not in tension with current observations and we suggest that GW190425 is a viable candidate of a neutron star$-$black hole merger event. Benefitting from the continual enhancement of the sensitivities of the advanced gravitational detectors and the increase of the number of the observatories, similar events are anticipated to be much more precisely measured in the future and the presence of black holes below the so-called mass gap will be unambiguously clarified. If confirmed, the mergers of neutron stars with (quickly rotating) low-mass black holes are likely important production sites of the heaviest r-process elements.
\end{abstract}

\section{introduction}
The neutron star (NS)$-$black hole (BH) binary systems, though not directly observed before, have been widely believed to exist in the universe \citep[see][and the references]{2010CQGra..27q3001A}. In addition to giving rise to strong gravitational-wave (GW) radiation, the NS$-$BH mergers can also produce electromagnetic transients such as short/long-short gamma-ray bursts (GRBs) and macronovae/kilonovae, as long as the merging neutron stars have been effectively tidally disrupted \citep[e.g.,][]{Narayan1992,1998ApJ...507L..59L,Piran2004,Metzger2019}.
In the absence of GW observations, the well-measured macronova/kilonova signals in the afterglow of some short/long-short GRBs in principle can shed valuable light on the merger nature \citep[e.g.,][]{Hotokezaka2013,2020arXiv200104474K}. Indeed, the NS$-$BH merger model has been adopted to well reproduce the luminous and relatively blue macronova/kilonova signal of the long-short GRB 060614 \citep{2015ApJ...811L..22J,2015NatCo...6.7323Y}. According to the macronova/kilonova modeling of a few events, the NS$-$BH merger rate was estimated to be $\sim {\rm a~few}~\times~100~{\rm Gpc^{-3}~yr^{-1}}$ and some BHs were speculated to have low masses \citep{2017ApJ...844L..22L}. These arguments are indirect and more solid evidence for the NS$-$BH mergers is highly needed. Such a purpose can be achieved in the GW observations.

The data of GW170817, the first neutron star merger event, strongly favor the binary neutron star (BNS) merger scenario \citep{2017PhRvL.119p1101A,2019PhRvX...9a1001A}. Though the NS$-$BH merger possibility has also been examined, the inferred masses of the involved BH and NS are not natural \citep{Coughlin2019,2019PhRvD.100f3021H}. Very recently, the LIGO/Virgo collaboration reported the detection of GW190425, the second neutron star merger event with a total gravitational mass of $M_{\rm tot}\approx 3.4M_\odot$ \citep{2020arXiv200101761T}. These authors concluded that such a massive binary most likely consists of a pair of NSs and has intriguing implications on the stellar evolution. The lack of the detection/identification of such massive binaries in the Galaxy, which is unlikely attributed to their quick merging after the birth \citep{arXiv:2001.04502}, motivates us to further examine the possible NS$-$BH merger origin of GW190425. Since the GW data alone are known to be unable to pin down the nature of the two compact objects \citep{2020arXiv200101761T}, our main purpose is to check whether the NS$-$BH merger hypothesis is in agreement with other data or not.

\section{The data analysis and the test of the BH hypothesis}
\subsection{The data analysis within the NS$-$BH merger scenario}
\label{sec:data-ana}

GW190425 was mainly detected by LIGO-Livingston (L1) on 2019 April 25 08:18:05.017 UTC \citep{2020arXiv200101761T}.
LIGO-Hanford (H1) was offline at the time. The signal-to-noise ratio (S/N) of the Advanced Virgo (V1) is low, but
it is consistent with the L1 data given the relative sensitivities of the detectors \citep{2020arXiv200101761T}.
To obtain the source parameters of the GW merger event, we apply the widely used Bayesian parameter inference method. Based on the work of \citet{2020arXiv200101761T}, we take the cleaned data spanning GPS time $(1240215303, 1240215511){\rm s}$ which are open access and available from the Gravitational Wave Open Science Center\footnote{\url{https://www.gw-openscience.org/eventapi/html/O3_Discovery_Papers/GW190425/v1/}} \citep{2015JPhCS.610a2021V}. Due to the low S/N of GW190425, systematic errors caused by the choice of waveform is negligible compared to the large statistical uncertainties. Therefore, we take the spin-aligned waveform template IMRPhenomDNRT \citep{2016PhRvD..93d4006H, 2016PhRvD..93d4007K, 2017PhRvD..96l1501D, 2019PhRvD..99b4029D} to analyze the data, and use the SEOBNRv4\_ROM \citep{2017PhRvD..95d4028B} with added tidal phase correction \citep{2017PhRvD..96l1501D} to check the result. We do not consider the calibration errors of the detector that will influence the sky localization but has little effect on mass measurements \citep{2016PhRvL.116x1102A}. For the noise power spectral density (PSD), we take the files from LIGO Document Control Center\footnote{\url{https://dcc.ligo.org/LIGO-P2000026/public}}. Then the single-detector log-likelihood can be constructed with the GW data $d(f)$, one-sided PSD $S_{\rm n}(f)$, and waveform model $h(\vec{\theta}_{\rm GW},f)$, which reads
\begin{equation}
\label{eq:Likelihood}
{\log L}(d|\vec{\theta}_{\rm GW}) = -2 \int_{f_{\rm min}}^{f_{\rm max}} \frac{|d(f) - h(\vec{\theta}_{\rm GW},f)|^2}{S_{\rm n}(f)} df + C,
\end{equation}
where we take $f_{\rm min}=19.4 {\rm Hz}$ and $f_{\rm max}=2048 {\rm Hz}$ following \citet{2020arXiv200101761T}. With the likelihood in hand, it is convenient to estimate the posterior probability distributions for the source model parameters using the Bayesian stochastic sampling software; we use the \textsc{PyCBC} Inference \citep{2019PASP..131b4503B} with the sampler \textsc{dynesty} \citep{2019arXiv190402180S} for our analysis and the \textsc{Bilby} \citep{2019ApJS..241...27A} with the sampler \textsc{PyMultiNest} \citep{2016ascl.soft06005B} for checking. To accelerate Nest sampling \citep{2004AIPC..735..395S}, we marginalize the likelihood over the coalescence phase \citep{2012PhRvD..85l2006A, 2019PhRvX...9a1001A, 2019EPJA...55...50R, 2019PASA...36...10T}. Thus the parameters of GW take the form $\vec{\theta}_{\rm GW} = \{\mathcal{M}, q, \chi_{\rm BH}, \chi_{\rm NS}, D_{\rm L}, \theta_{\rm jn}, {\rm R.A.}, {\rm decl}, t_{\rm c}, \Psi, \Lambda_{\rm BH}, \Lambda_{\rm NS}\}$, where $\mathcal{M}$, $q$, $\chi_{\rm BH}$($\chi_{\rm NS}$), $D_{\rm L}$, $\theta_{\rm jn}$, ${\rm R.A.}$, ${\rm decl}$, $t_{\rm c}$, $\Psi$, and $\Lambda_{\rm BH}(\Lambda_{\rm NS})$ are chirp mass, mass ratio, aligned spins, luminosity distance, inclination angle, right ascension, declination, geocentric GPS time of the merger, polarization of GW, and dimensionless tidal deformabilities, respectively.

For the NS$-$BH merger scenario, we set the prior of $q=M_{\rm NS}/M_{\rm BH}$ to a uniform distribution in the range of $(0.2,1)$ (we also set the prior of $q^{-1}$ into a log-uniform distribution in the range of $(1, 5)$, and find that the results just change slightly), and take $\Lambda_{\rm BH}=0$, while $\Lambda_{\rm NS}$ is assumed to lie in a wide range $(0, 10,000)$ uniformly. Meanwhile, we give a low-spin prior for the component of the spin aligned with the orbital angular momentum of NS $|\chi_{\rm NS}|<0.05$ \citep{2020arXiv200101761T}, and a much broader prior for that of BH, i.e., $|\chi_{\rm BH}|<0.998$ \citep{Thorne1974}. The chirp mass $\mathcal{M}$ is uniformly distributed in $(1.42,2.6) M_{\odot}$ \citep[in the detector frame;][]{2020arXiv200101761T}, the luminosity distance is uniform in comoving volume bounded in $(1,500){\rm Mpc}$, and other parameters are all uniformly distributed in their domains. Additionally, we assume the source frame mass $M_{\rm BH}>2.04M_\odot$, i.e., it is above the 1$\sigma$ lower limit on the mass of PSR J0740+6620 \citep[][the uniform rotation of this pulsar can enhance the gravitational mass by $\sim 0.01M_\odot$, which has been corrected here]{2019NatAs.tmp..439C}, and assume $M_{\rm NS}>1.0M_{\odot}$, as widely anticipated in the literature \citep[see][for a review]{Lattimer2012}.

\begin{figure}[!ht]
\begin{center}
\includegraphics[width=1.0\textwidth]{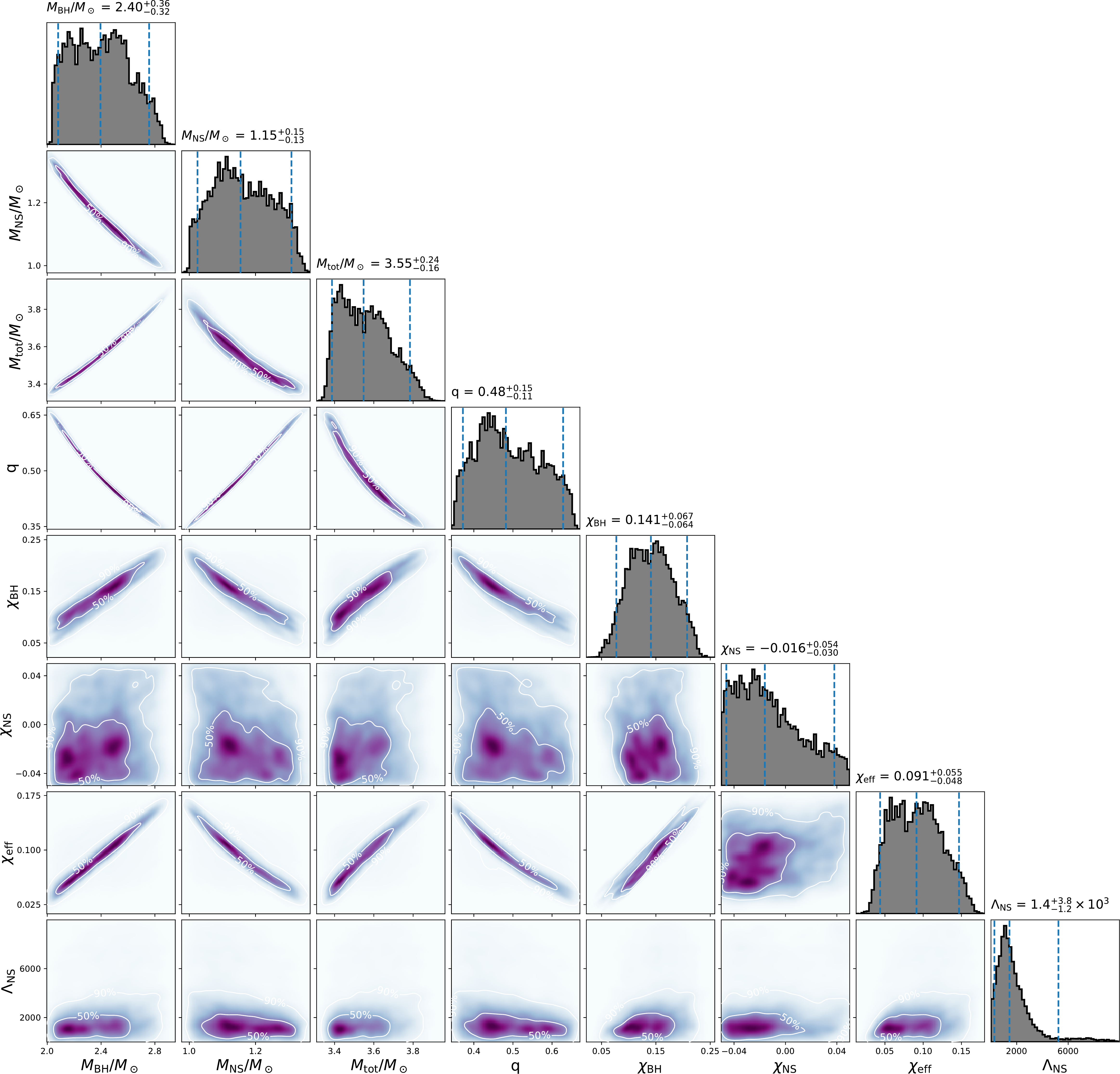}
\end{center}
\caption{Posterior distributions of the physical parameters, including the source frame masses of the two compact objects $(M_{\rm BH},~M_{\rm NS})$, the source frame total mass $M_{\rm tot}$, the mass ratio $q$, the dimensionless spins $(\chi_{\rm BH},~\chi_{\rm NS})$, the effective spin parameter $\chi_{\rm eff}$, and the dimensionless tidal deformability of the neutron star $\Lambda_{\rm NS}$. The error bars are all for the 90\% credible level.}
\label{fig:MCMC}
\end{figure}

Fig.\ref{fig:MCMC} presents the main results of our data analysis (a crosscheck of our codes can be found in the Appendix \ref{appdx:high}).
It contains the 2D density plots and the marginal distributions of some intrinsic parameters and their combinations. For the BH component we have $M_{\rm BH} \in (2.08, 2.76) M_\odot$ (i.e., the primary mass; in this work the ranges of the parameters represent the 90\% credible intervals). While for the mass of the NS component, we have $M_{\rm NS} \in (1.02, 1.30) M_\odot$. The total gravitational mass and the mass ratio of the binary are $3.55^{+0.24}_{-0.16} M_\odot$ and $q \in (0.37, 0.63)$, respectively. The dimensionless aligned spin of the BH is $\chi_{\rm BH}=0.141^{+0.067}_{-0.064}$. As for the NS, we have $\chi_{\rm NS}=-0.016^{+0.054}_{-0.030}$. The inferred mass and spin of the NS component are consistent with the observations of the Galactic binary systems. Besides, the effective spin, defined as $\chi_{\rm eff}=(M_{\rm BH}\chi_{\rm BH}+M_{\rm NS}\chi_{\rm NS})/M_{\rm tot}$, is found to be $0.091^{+0.055}_{-0.048}$. We do not show the results of other extrinsic parameters that are just poorly constrained because of the absence of the electromagnetic counterparts and the nondetection by H1 (the S/N of V1 is very low).

\begin{figure}[!ht]
\centering
\includegraphics[width=0.45\columnwidth]{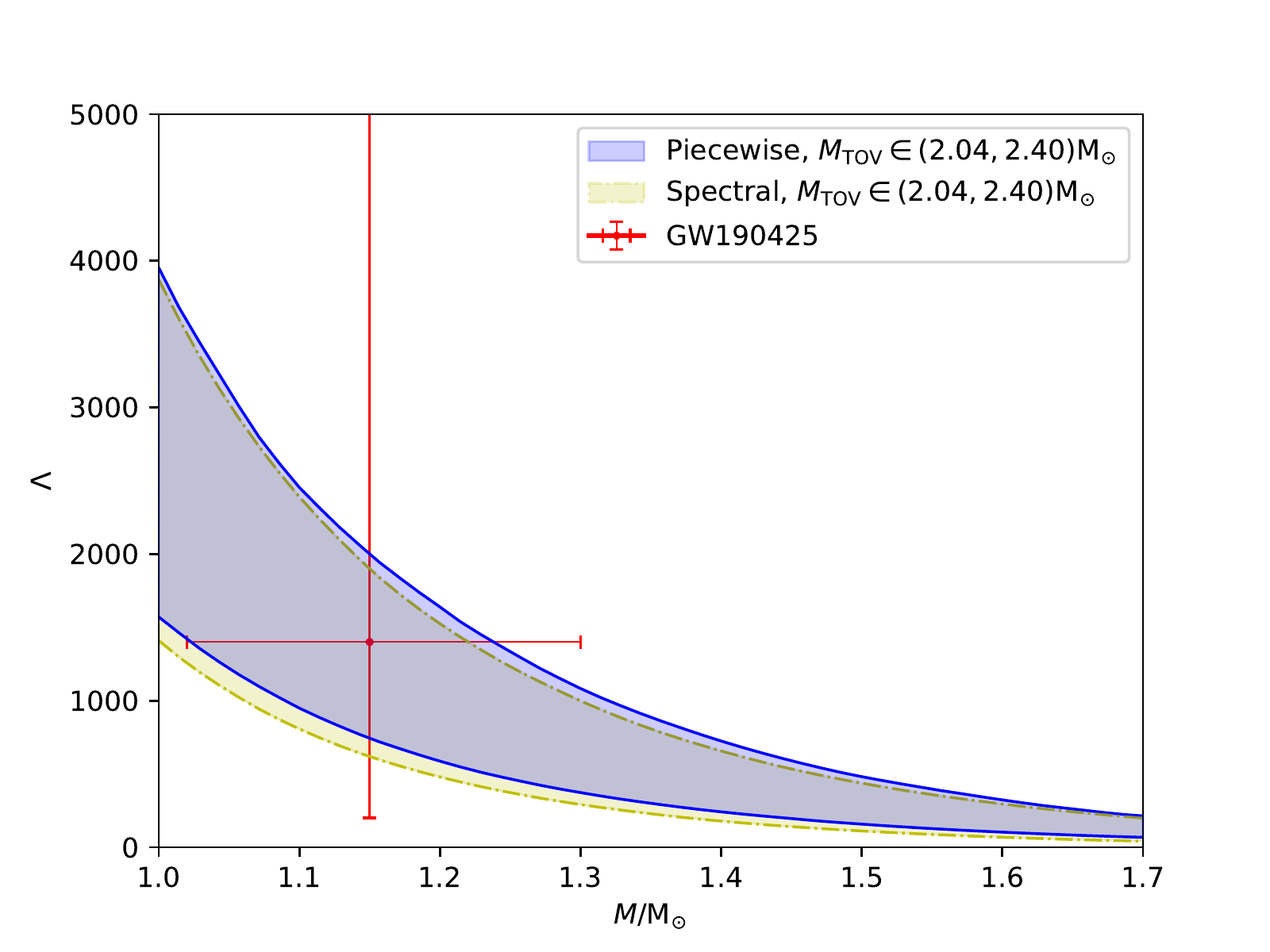}
\includegraphics[width=0.45\columnwidth]{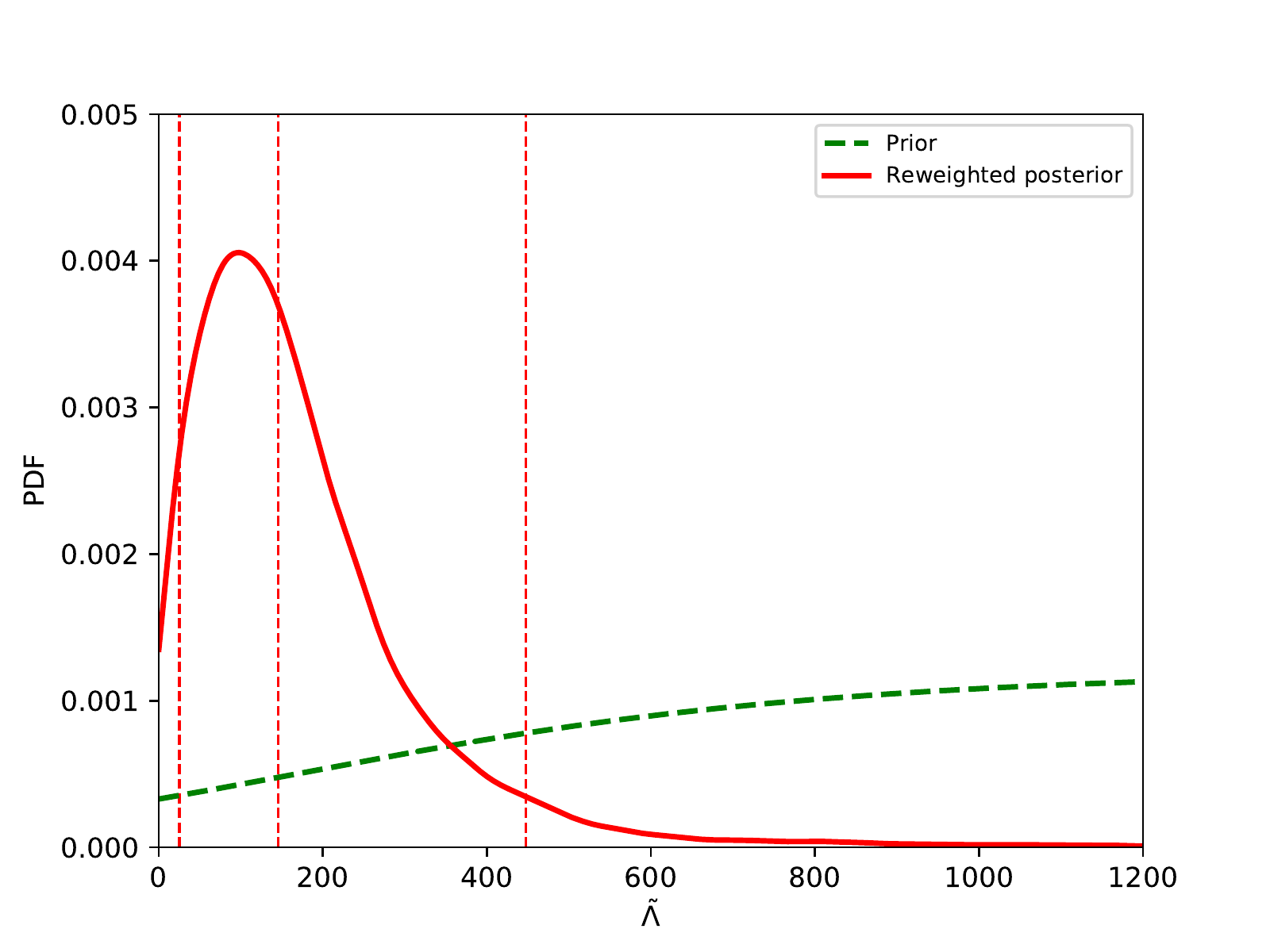}
\caption{Left panel (a): the inferred $M$ and $\Lambda$ of the neutron star component of GW190425 at the 90\% credible level. The shaded regions represent the jointed constraints set by GW170817, PSR J0030+0451 and the nuclear data, that are adopted from \citet{2019arXiv191207467J}. Right panel (b): posterior probability density functions (PDFs) of the combined tidal parameter $\tilde{\Lambda}$ of the NS component which is re-weighted by the prior. The vertical dashed lines denote the 90\% credible interval.}
\label{fig:ld}
\end{figure}

The dimensionless tidal deformability of the NS is constrained to $\Lambda_{\rm NS}=(2/3)k_{2}[(c^2/G)(R_{\rm NS}/M_{\rm NS})]^5=1.40^{+3.80}_{-1.20}\times10^{3}$, where $k_{2}$ is the tidal Love number \citep{2008ApJ...677.1216H, 2008PhRvD..77b1502F}, $c$ is the speed of light in vacuum, $G$ is the gravitational constant, and $R_{\rm NS}$ is the radius of NS, respectively. Fig.\ref{fig:ld}(a) shows the result of the tidal deformability and the mass of NS. In comparison to GW170817, the signal of GW190425 has a considerably lower S/N (partly attributing to the non-observation of the H1 detector) and the constraint on $\Lambda$ is looser \citep{2020arXiv200101761T}. Anyhow, the resulting $\Lambda$ is consistent with the joint constraints set by GW170817, PSR J0030+0451 and the nuclear data \citep{2019arXiv191207467J}. Note that for the signal with a low S/N, the inferred $\Lambda$ is likely biased to a higher value \citep{Han2020}. Besides, as shown in Fig.\ref{fig:ld}(b) the combined tidal parameter $\tilde{\Lambda}$ is constrained to $\tilde{\Lambda}=171^{+378}_{-141}$, which is given by $\tilde{\Lambda}=16(M_{\rm NS}+12M_{\rm BH})M_{\rm NS}^4\Lambda_{\rm NS}/(13{M_{\rm tot}^5})$ for $\Lambda_{\rm BH}=0$ \citep{2014PhRvD..89j3012W}.

\subsection{Testing the validity of the BH hypothesis}
In principle, the massive BNS merger model can be distinguished from the NS$-$BH merger model if the macronova/kilonova emission has been well monitored \citep{2020arXiv200104474K}. However, GW190425 was only poorly localized and the partial volume covered by macronova/kilonova observations just span up to about $40\%$ \citep[][for AT 2017gfo-like macronova/kilonova]{2019ApJ...880L...4H}, which are insufficient to pin down the merger scenario. In this subsection we concentrate on the possibility that the heavy component is a BH.

So far, thanks to the long time radio observations of the massive pulsars, the record of observed maximum mass of NS was broken over and over, e.g., from $2.01\pm0.04M_{\odot}$ \citep[PSR J0348+0432,][]{2013Sci...340..448A} to $2.14^{+0.10}_{-0.09}M_{\odot}$ \citep[PSR J0740+6620,][]{2019NatAs.tmp..439C}. Though the maximum mass of a nonrotating neutron star (i.e., $M_{\rm TOV}$) still remains unknown, one can statistically estimate the probability distribution of NS's maximum mass ($M_{\rm max}$, which can be approximated by $M_{\rm TOV}$ since the rotation of all these NSs are slow that do not effectively enhance the gravitational mass) using the mass measurements of dozens of NSs. Such a study has been recently carried out by \citet{2018MNRAS.478.1377A}, where the authors have found strong evidence for the presence of a maximum mass cutoff. At that moment the mass of PSR J0740+6620 was unavailable, and their sample included a few massive neutron stars such as PSR J0348+0432, Vela X-1 ($2.12\pm0.16M_\odot$), and PSR J1748-2021B ($2.74\pm0.21M_\odot$, supposing the observed periastron advance rate of this binary system is purely relativistic).
\begin{figure}[!ht]
\begin{center}
\includegraphics[width=0.6\textwidth]{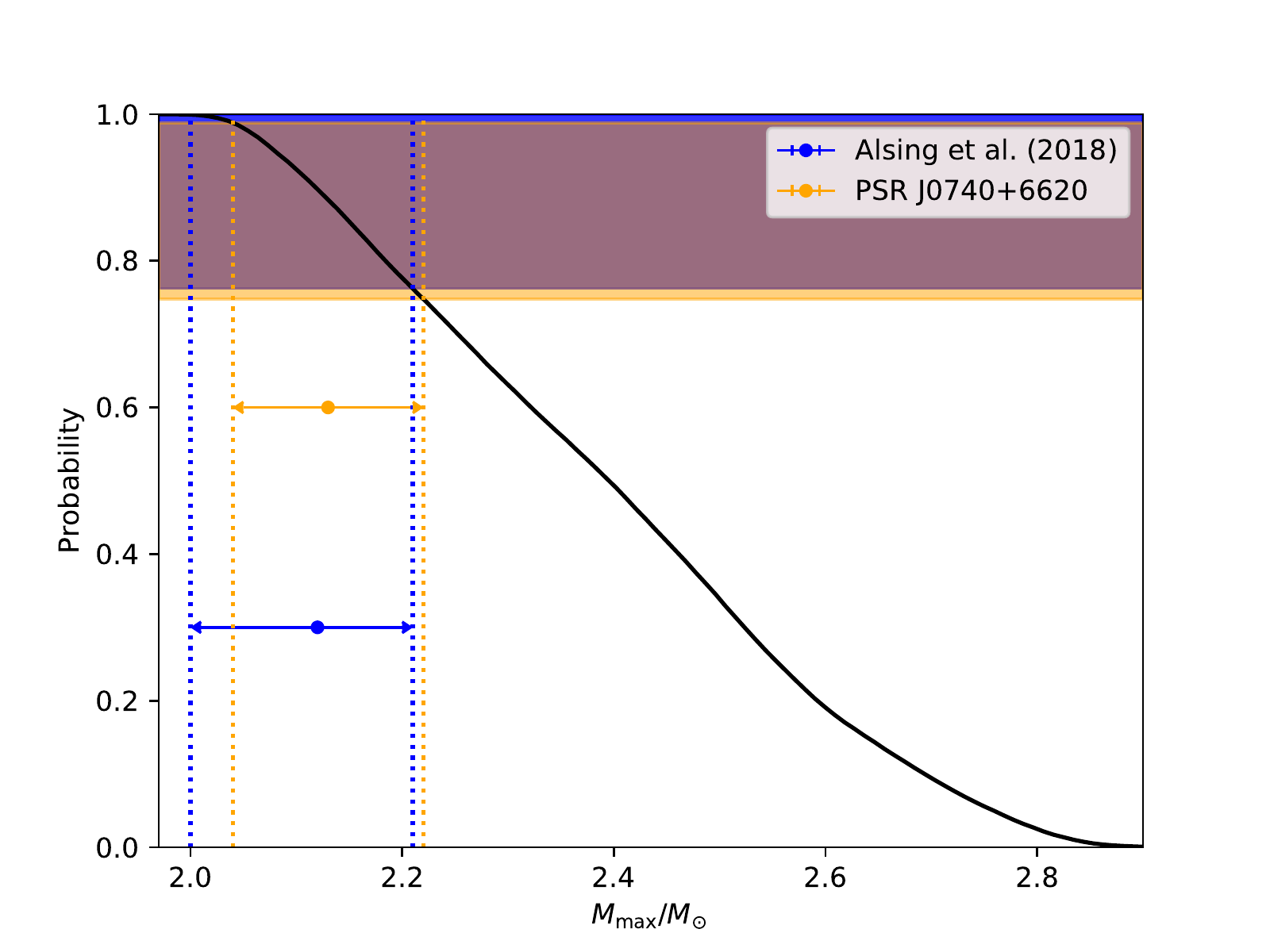}
\end{center}
\caption{Conditional probability for the heavier component's mass being above a given $M_{\rm max}$. The blue lines represent the 68\% confidence region of the maximum mass cutoff of NSs \citep{2018MNRAS.478.1377A}, while the orange lines show the 68\% credible interval of the gravitational mass of PSR J0740+6620 \citep{2019NatAs.tmp..439C}. The shaded areas show the plausible regions of the probability ${\cal P}(M_{\rm BH}>M_{\rm max}|M_{\rm max})$ for these two cases. Clearly, the heavy component of GW190425 is consistent with being a black hole.}
\label{fig:BH-prob}
\end{figure}
Here we evaluate the possibility of the heavy component of GW190425 being a BH with the condition of $M_{\rm BH}>M_{\rm max}$.
Using the posterior of $\mathcal{M}$, $q$, and $D_{\rm L}$ obtained in Sec.\ref{sec:data-ana}, it is straightforward to calculate the posterior probability distribution of the heavier component's mass with
\begin{equation}
\label{eq:mcqtom12}
M_{\rm BH}=\frac{q^{-3/5}(1+q)^{1/5}\mathcal{M}}{1+z(D_{\rm L})},
\end{equation}
where $z$ is the cosmic redshift transformed from luminosity distance $D_{\rm L}$ assuming the ${\rm \Lambda CDM}$ cosmology \citep{2016A&A...594A..13P}.
Then we integrate the probability distribution of the heavier component's mass $P(M_{\rm BH})$ with the condition of $M_{\rm BH}>M_{\rm max}$ through
\begin{equation}
\label{eq:possibility}
{\cal P}(M_{\rm BH}>M_{\rm max}|M_{\rm max})= \int_{M_{\rm max}}^{\infty} P(M_{\rm BH}) \, \mathrm{d} M_{\rm BH}\, ,
\end{equation}
and the results are shown in Fig.\ref{fig:BH-prob}, which indicate that our NS$-$BH merger assumption is self-consistent.
\citet{2020arXiv200101761T} found a rate of GW190425-like events of $460^{+1050}_{-390}~{\rm Gpc^{-3}~yr^{-1}}$, which is comparable to that suggested in \citet[][see Tab.1 therein]{2017ApJ...844L..22L} but may be hard to achieve in some population synthesis calculations \citep[e.g.,][in which a local NS$-$BH merger rate of $\leq 150~{\rm Gpc^{-3}~yr^{-1}}$ has been suggested]{Cote2017}, implying that some new NS$-$BH binary formation channels may be present.

\section{Discussion}
In this work we have examined the possible NS$-$BH origin of GW190425, the second neutron star merger event detected by the Advanced LIGO/Virgo detectors. In such a specific scenario, the GW data favor a BH (NS) mass of $2.40^{+0.36}_{-0.32}M_\odot$ ($1.15^{+0.15}_{-0.13}M_\odot$) and an aligned spin (a dimensionless tidal deformability) of $0.141^{+0.067}_{-0.064}$ ($1.4^{+3.8}_{-1.2}\times 10^{3}$). The inferred parameters are not in tension with current observations and we suggest that GW190425 is a viable candidate of an NS$-$BH merger event. This is different from the case of GW170817, for which the NS$-$BH modeling yields unnatural masses of the objects \citep{Coughlin2019,2019PhRvD.100f3021H}. Therefore, GW190425 may be the first detected NS$-$BH merger event. The current data, however, are insufficient to disfavor the double neutron star merger origin \citep{2020arXiv200101761T} because of the low S/N of the signal and the nondetection of the electromagnetic counterparts. GW190425 was just detected by the LIGO-Livingston Observatory and the Virgo Observatory, but not the LIGO-Hanford Observatory. Currently, the sensitivity of the Virgo Observatory is considerably lower than that of LIGO-Livingston and LIGO-Hanford Observatories, and hence can not contribute significantly to improving the S/N of the signal (anyhow, the presence of a low-S/N signal in the Virgo Observatory provides a valuable verification). Together with the information reported in https://gracedb.ligo.org/superevents/public/O3/, the duty cycle for each detector of current aLIGO is an important issue. The situation will change substantially in the near future. The Kamioka Gravitational Wave Detector (KAGRA) will join the O3 run of the Advanced LIGO/Virgo network in 2020. The sensitivities of Virgo and KAGRA will be enhanced by a factor of a few in the upcoming O4 run. LIGO-India is anticipated to join in 2025. Therefore, for GW190425-like events taking place in O4 and later runs of LIGO/Virgo/KAGRA, the S/N would be higher by a factor of $\sim 2-6$, benefitting from the enhancement of the sensitivity of the advanced gravitational detectors and the increase of the number of the observatories. With such a high S/N, the GW data will provide much more accurate classification of the compact objects. The joint observation of multiple detectors will improve the localization of the mergers considerably, which is very helpful to catch the macronova/kilonova radiation (and the off-axis afterglow emission), with which the nature of merger can be further revealed.

If the NS$-$BH merger origin of GW190425-like events has been confirmed, there are some interesting implications: (i) There exists low-mass BH below the so-called mass gap, which favors the formation of BH with a continual mass distribution rather than a gap, as suggested for instance in \citet{2019ApJ...870....1E} and \citet{2020MNRAS4912715B}. The continual distribution of the BH masses may lead to the misidentification of a binary BH system with light components (e.g., $\sim 3M_{\odot}$) into NS$-$BH systems, which consequently brings difficulty for constructing the BH mass function of such systems \citep{2018ApJ...856..110Y,2020arXiv200207573T}. (ii) The NS$-$quickly rotating low-mass black hole mergers could eject massive subrelativistic neutron-rich outflow \citep{Lattimer1974}. In comparison to the BNS merger scenario, very heavy r-process elements are likely easier to form because for the former the huge amount of neutrino emission from the pre-collapse massive neutron stars will make the subrelativistic ejecta less neutron-rich. Therefore, together with a high rate, such mergers can be important sites of the heaviest r-process nucleosynthesis \citep{2015NatCo...6.7323Y,Jin2016,2019MNRAS.487.1745W}.

\section*{Acknowledgments}
We thank the anonymous referee and Dr. G. Ashton for the helpful suggestions/comments.
This work was supported in part by NSFC under grants of No. 11525313 (i.e., Funds for Distinguished Young Scholars), No. 11921003, No. 11773078, and No. 11933010, the Funds for Distinguished Young Scholars of Jiangsu Province (No. BK20180050), the Chinese Academy of Sciences via the Strategic Priority Research Program (grant No. XDB23040000), Key Research Program of Frontier Sciences (No. QYZDJ-SSW-SYS024). This research has made use of data and software obtained from the Gravitational Wave Open Science Center; \url{https://www.gw-openscience.org}, a service of LIGO Laboratory, the LIGO Scientific Collaboration and the Virgo Collaboration. LIGO is funded by the U.S. National Science Foundation. Virgo is funded by the French Centre National de Recherche Scientifique (CNRS), the Italian Istituto Nazionale della Fisica Nucleare (INFN) and the Dutch Nikhef, with contributions by Polish and Hungarian institutes.\\

\software{Bilby \citep[version 0.5.5, ascl:1901.011, \url{https://git.ligo.org/lscsoft/bilby/}]{2019ascl.soft01011A}, Dynesty \citep[version 1.0, ascl:1809.013, \url{https://github.com/joshspeagle/dynesty/tree/v1.0.0}]{2018ascl.soft09013S}, PyCBC \citep[version 1.13.6, ascl:1805.030, \url{http://doi.org/10.5281/zenodo.3265452}]{2018ascl.soft05030T}, PyMultiNest \citep[version 2.6, ascl:1606.005, \url{https://github.com/JohannesBuchner/PyMultiNest}]{2016ascl.soft06005B}}

\appendix
\section{The crosscheck of our data analysis}
\label{appdx:high}

\begin{figure}[!ht]
\begin{center}
\includegraphics[width=1.0\textwidth]{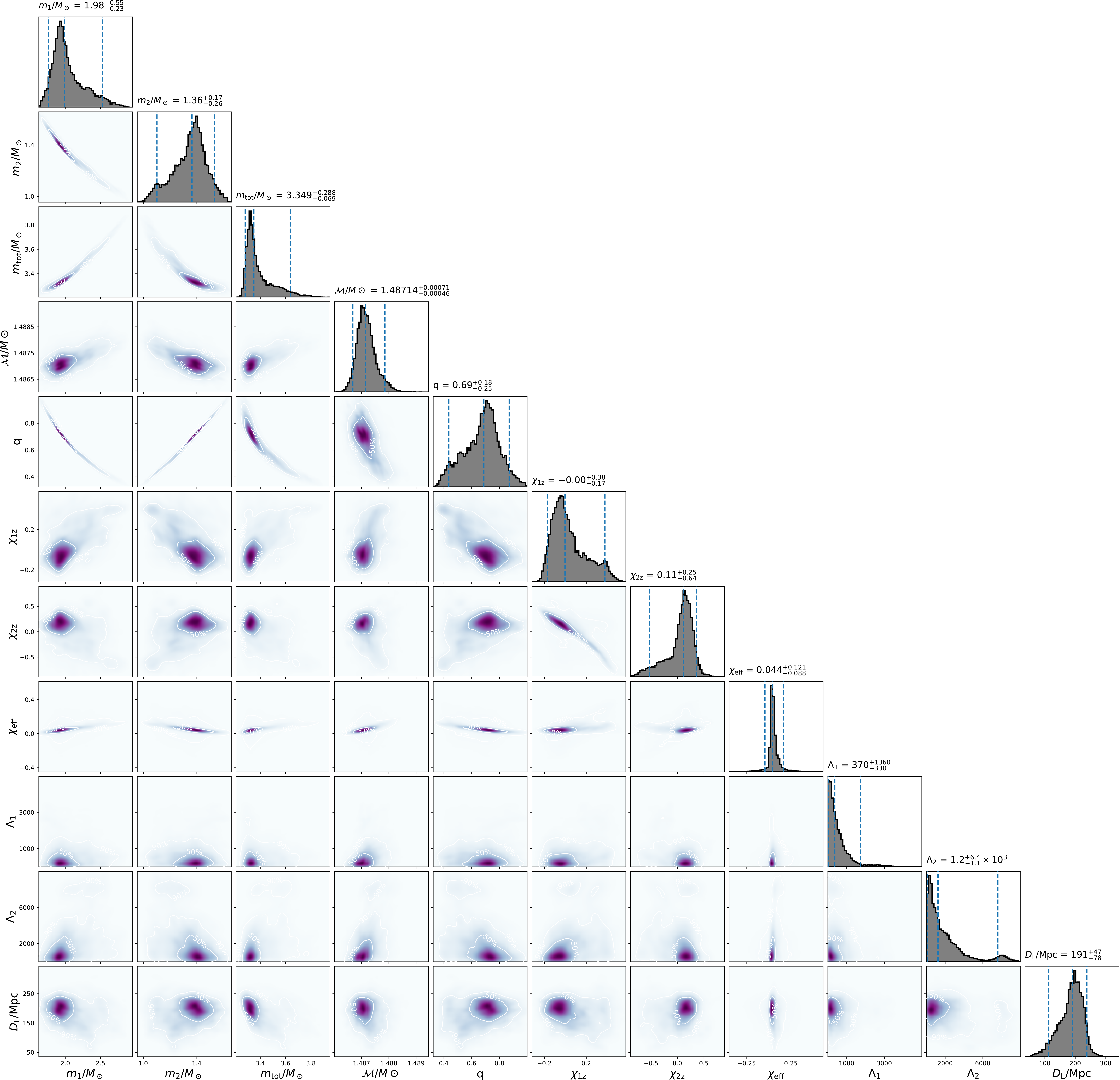}
\end{center}
\caption{Posterior distributions of the physical parameters with the same assumptions of \citet[][the case of high spin prior]{2020arXiv200101761T}, including the source frame masses of the two compact objects $(m_1,~m_2)$, the source frame total mass $M_{\rm tot}$, the detector frame chirp mass $\mathcal{M}$, the mass ratio $q$, the dimensionless spins $(\chi_{\rm 1z},~\chi_{\rm 2z})$, the effective spin parameter $\chi_{\rm eff}$, the dimensionless tidal deformabilitis $(\Lambda_1,~\Lambda_2)$, and the luminosity distance $D_L$. The error bars are all for the 90\% credible level. The results are consistent with that reported in \citet{2020arXiv200101761T}.}
\label{fig:high}
\end{figure}

As a crosscheck, here we adopt the assumptions of \citet[][the case of high spin prior]{2020arXiv200101761T} and then reanalyze the GW data of GW190425. Different from our approach in the main text, \citet{2020arXiv200101761T} set constraints of $(1.0, 5.31)~M_\odot$ on the mass range and the dimensionless spin $|\chi|<0.89$ for both compact objects. The results are presented in Fig.\ref{fig:high}, where we show the 2D density plots and the marginal distributions of some parameters and their combinations. The source frame uncertainty ranges of $m_1,~m_2,~M_{\rm tot}$ reported in \citet{2020arXiv200101761T} are $(1.61, 2.52)~M_\odot$, $(1.12, 1.68)~M_\odot$, and $3.4^{+0.3}_{-0.1}~M_\odot$, respectively. While in our analysis, they are $(1.75, 2.53)~M_\odot$, $(1.10, 1.53)~M_\odot$, and $3.35^{+0.29}_{-0.08}~M_\odot$, respectively. Our combined tidal parameter is $\tilde{\Lambda}=715^{+1418}_{-538}$, while \citet{2020arXiv200101761T} found a $\tilde{\Lambda}<1900$. These two groups of results are similar, which in turn validates our analysis. Anyhow, our results (for example the $q$ range) are not exactly the same as that of \citet{2020arXiv200101761T}, which might be caused by our simplifications of the approach (for instance, we have ignored the calibration errors of the detector) and/or by the different analysis procedures (We perform the parameter estimation by using the \textsc{PyCBC} Inference and the \textsc{dynesty} sampler, while these authors used the \textsc{LALInference} \citep{2015PhRvD..91d2003V} within \textsc{LALSuite} \citep{lalsuite}. Besides, a uniform rather than a log-uniform prior for $q$ is adopted in this work, which may also shape the resulting $q$ distribution in view of the low SNR of the current signal).

\end{document}